\documentclass[aps,prl,floats,twocolumn,showpacs]{revtex4-1}

\usepackage{graphicx}

\usepackage{amsmath}
\usepackage{hyperref}
\usepackage[dvips]{color}
\usepackage{amssymb}

\usepackage{bm}

\def\up{\uparrow}
\def\down{\downarrow}
\def\ud{\uparrow\downarrow}
\def\du{\downarrow\uparrow}
\def\uu{\up\up}
\def\dd{\down\down}
\def\Tr{\mbox{Tr}\,}
\def\Im{\mbox{Im}}
\def\Re{\mbox{Re}}

\begin{document}

\title{Charge Resistance in a Majorana RC circuit}

\author{Anatoly Golub and Eytan Grosfeld}
\affiliation{Department of Physics, Ben-Gurion University of the Negev Beer-Sheva, Israel}
\date{\today}

\begin{abstract}
We investigate the dynamical charge response in a ``Majorana Coulomb box'' realized by two Majorana bound states hosted at the ends of a mesoscopic topological superconductor. One side of the wire is coupled to a normal lead and low frequency gate voltage is applied to the system. There is no dc-current; the system can be considered as an RC quantum circuit. We calculate the effective capacitance and charge relaxation resistance. The latter is in agreement with the Korringa-Shiba formula where, however, the charge relaxation resistance is equal to $h/2e^2$. This value corresponds to the strong Coulomb blockade limit described by a resonant model formulated by Fu [PRL {\bf 104}, 056402 (2010)]. We also performed direct calculations using the latter model and defined its parameters by direct comparison with our perturbation theory results.
\end{abstract}


\maketitle


{\it Introduction}.---
The search for Majorana fermions (MFs) and Majorana bound states (MBSs) in condensed matter systems is currently under way, driven both by their unique physical properties and by their potential applications in topological quantum computation \cite{alicea,nayak}. There are several promising systems predicted to host MFs, including: spin-triplet $p$-wave superconductors, topological insulator/superconductor heterostructures and semiconductor/superconductor junctions. Of particular interest is the case where a semiconducting nano-wire with strong spin-orbit coupling is fabricated on top of a superconductor. The nano-wire can be tuned into a topological phase, supporting MBSs at its two ends, by applying a Zeeman field onto it and tuning the chemical potential into the Zeeman gap \cite{sarma,oreg}. For this system the zero-bias conductance peak, recently observed in the tunneling current from a normal lead, would be considered as evidence for the presence of MBSs \cite{sarma,oreg,kouwn}. Signatures of  MFs in topological Josephson junctions \cite{fukane,grosfeld}, which consist of two superconducting leads coupled through a three-dimensional topological insulator, have also been reported \cite{gordon}.

In a typical experimental setup for realizing MFs the host topological superconductor (TS) is placed in proximity to a top gate. An intriguing and experimentally relevant question is whether MFs, being inherently charge-neutral objects, can participate in the charge response of the system to the nearby gate. In this Letter we study the effect of a small time-dependent oscillatory component of the gate voltage on a mesoscopic system supporting MBSs which is coupled to a single-electron bath. We show that, remarkably, the MBSs play a dominant role in the response to the oscillating gate voltage; and that, in particular, they give rise to a sub-gap dissipative charge response. More specifically, the low frequency charge response of the system is encapsulated in two quantities: an effective capacitance and a dissipative charge relaxation resistance. We demonstrate that both are sensitive to the presence of MBSs, and contain unique signatures of the latter (see Fig.~2, Eq.~(\ref{rr}), Eq.~(\ref{ii}) and the discussion following them). Retaining the Coulomb interaction is crucial for correctly identifying these quantities. We expect that these concepts will play an important role in future experiments attempting to measure signatures of MFs embedded in mesoscopic settings.

The linear charge response to gate voltage oscillations for the setup of a quantum dot connected to a single lead has attracted considerable interest \cite{gabelli,buttiker,mora, mora1,mora2,rosa,rodionov}. In the coherent tunneling regime the charge response defines a mesoscopic capacitance $C_0$ and a charge relaxation resistance $R_q$ according to
\begin{equation}\label{q}
    \frac{Q(\omega)}{V_g (\omega)}=C_0 (1+i\omega C_0 R_q),
\end{equation}
where $Q$ is the charge on the dot and $V_g$ is the gate voltage. The capacitance $C_0$ describes the
static charging of the dot. It is generally different \cite{grabert,glazman} from
the geometrical capacitance $C$ and depends strongly on the lead-dot tunneling width $\Gamma$. Energy dissipation is associated with a time scale $C_0 R_q$, where $R_q$ is known as the charge relaxation resistance. It assumes the quantized value $R_q= h/(2e^2)$ which, unlike the classical limit, for a single channel junction does not depend on the transmission. This prediction remains valid also when effects of interaction are taken into consideration \cite{mora,mora1}. The quantum impurity models that have been studied so far to analyze the charge response include the Anderson model \cite{mora,rosa} and the model for large metallic dots, described by  the Ambegaokar-Eckern-Schon effective action \cite{rodionov}.

\begin{figure} [!ht]
\centering
\includegraphics[width=0.45 \textwidth]{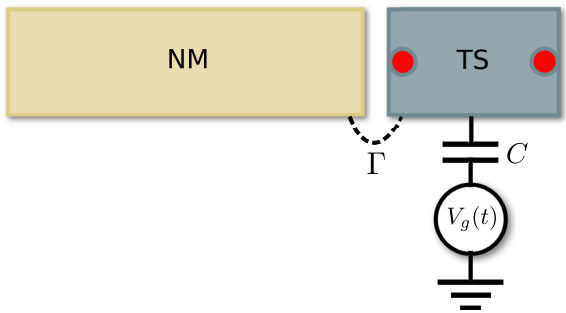}
\caption {\footnotesize{A schematic of a quantum RC circuit with Majorana modes (represented as two red dots). A topological superconductor which hosts two MBSs at its ends is coupled via a tunnel contact (with hybridization tunneling width $\Gamma$) to a normal metal lead. The dissipative current through the tunneling contact is induced by the ac gate voltage $V_g(t)$. The charging energy strongly influences this current and the effective capacitance $C_0$.
}}
\label{Fig1}
\end{figure}

We now proceed to analyze the linear charge response in a different quantum impurity model, presented in Fig.~1, where a mesoscopic TS wire supporting two MBSs at its ends \cite{alicea,oreg,sarma} takes the role of the quantum dot. The wire is coupled to a top gate and, on one of its ends, to a normal metal (NM) lead. The potential of the normal metal lead is taken equal to zero $(V=0)$, while the potential of the gate  $V_g(t)$ includes a small time-dependent perturbation. The Coulomb interaction on the dot is described via the non-zero charging energy $E_c=e^2/2C$ where $C$ is the geometrical capacitance. The presence of a charging energy introduces dynamics for the superconductor phase which, combined with the coupling to the lead, induces phase fluctuations on the dot; therefore, it influences the dissipation. More specifically, the two Majorana fermions at the ends of the wire $\gamma_1$ and $\gamma_2$ (index 1 denotes the left side which is taken to interact with the  metallic lead), with anti-commutation relations $\{\gamma_i,\gamma_j\}=\delta_{ij}$, introduce the auxiliary fermion $d=(\gamma_1+i\gamma_2)/\sqrt{2}$.
The number operator $\hat{n}_d=d^{\dagger}d$ defines the occupation of this fermionic level. The two degenerate ground states with even or odd number of electrons are associated with different electron parities.  This fixes the occupation of the fermionic level according to $2\hat{n}_d-1=(-1)^{n+1}$ where $n$ is the electron number. Therefore, the MF response gets intermixed with the charge (or  equivalently its dual phase) dynamics \cite{fu}.

To calculate the charge response, we write the charge on the TS wire via the following formula
\begin{equation}\label{cc}
   Q(\omega)=e^2 K(\omega)V_g(\omega),
\end{equation}
which is expressed in terms of the retarded response function $K(t-t')=i\theta(t-t')\left\langle[\hat{n}_d(t),\hat{n}_d(t')]\right\rangle$, describing the charge fluctuations of the fermionic level at equilibrium (non-equilibrium response is also of interest \cite{baruch}  but is beyond the scope of this work). Here $\theta(t)$ is the step function: $\theta(t)=1$ if $t>0$ and is equal to zero if $t<0$.
Below we  proceed similarly to the approach presented in \cite{egger}  and construct a general Keldysh functional for the interacting problem using the dual phase representation for the Coulomb charging term.


{\it The Hamiltonian and the action}.---
We now derive the Keldysh generating functional in the presence of an external source term, which later will be used to calculate correlation functions. Our starting point is the Hamiltonian: The Majorana Coulomb box, lead and tunneling Hamiltonians are respectively \cite{egger,fu}
\begin{eqnarray}
  \nonumber & H_c=E_c\left(2\hat{N}+\hat{n}_d-n_0\right)^2,\,\, H_L=\sum_k \epsilon_k c_k^{\dagger}c_k,\\
  \label{Hc} & H_{t}=\lambda c^\dagger(0)\left(d+d^\dagger e^{-2i\phi}\right)+\mbox{h.c.},\label{hd}
\end{eqnarray}
where $n_0$ is tunable by the gate voltage $V_g$ and $\hat{N}$ is the number operator for Cooper-pairs, conjugate to the superconducting condensate phase $2\phi$. The lead Hamiltonian $H_L$ is expressed in terms of quasiparticle energies $\epsilon_k$ and electron operators $c_k, c^{\dagger}_k$. The operators $c(x),c^\dagger(x)$ are the Fourier transforms of the latter. By applying the Hubbard-Stratonovich transformation, the quadratic term in $\hat N$ is replaced by a gaussian form in $\phi$ with additional functional integration on the phases $\phi=(\phi_1,\phi_2)$ (where the phase $\phi$ is half of the condensate phase) \cite{nazarov,zaikin}. The total Lagrangian and, specifically, the Majorana sector of the Lagrangian, both depend on these phases. This directly shows that the charge transport which is described by the phase variables strongly influences the Majorana dynamics \cite{fu,egger}. The action of the dot takes the form
\begin{eqnarray}
          & S_c = \int dt  \left(\frac{\dot{\phi}^2} {4E_c} +n_0 \dot{\phi}+L_M\right)\sigma_z,\\
          & L_M =  \bar{d}\left(i\partial_t-\dot{\chi}\right)d.\label{lm}
        \end{eqnarray}
Here $\dot{\chi}=\dot{\phi}-\alpha$ where $\alpha$ is a source field. In the above formulas and throughout this Letter the Pauli matrices $\sigma_0,\sigma_x,\sigma_y,\sigma_z$  act within the Keldysh space. Accordingly all fields and sources have two components (denoted $1,2$).

It is convenient to gauge out the total field $\chi$ (including the source components) from the action $L_M$ by applying the transformation $d\rightarrow d e^{-i\chi}$. In this gauge the tunneling part of the action includes a phase factor
\begin{equation}\label{ht}
    S_t=-\lambda\int dt \bar{c}(0)\sigma_z\gamma_1 e^{-i\chi}+h.c,
\end{equation}
while the pure Majorana part becomes independent of $\chi$: $L_M=\frac{1}{2} \gamma^T[\sigma_z G^{-1}]\gamma$ where $G$ is the Majorana Green function (GF). Following a Keldysh rotation this GF acquires the form \cite{golub}
\[G=\left(
\begin{array}{cc}
G^K & G^R \\
G^A & 0 \\
\end{array}\right).
\]
Each element of this matrix is itself a matrix in Majorana basis. Below we use only the $11$-element of the latter matrices, being,
\begin{eqnarray}
    && G^R_{11}(\epsilon)=\frac{\epsilon}{(\epsilon+i\delta)^2-\nu_0^2}, \nonumber\\
    && G^K_{11}(\epsilon)=-i\pi \tanh\left(\frac{\epsilon}{2T}\right)[\delta(\epsilon-\nu_0)+\delta(\epsilon+\nu_0)].
\end{eqnarray}
Here in addition to the terms in Eq.~(\ref{lm}) we included a coupling energy $\nu_0$ between the two MBSs.

Integrating  out the lead operators and performing the Keldysh rotations of basis $( \gamma^{T}{'}, \gamma')$=($\gamma^{T}\sigma_{z} L^{-1}, L\sigma_{z}\gamma $), where $L=\label{L}\frac{1}{\sqrt 2}(\sigma_0-i\sigma_y)$, we obtain the effective action as function of the phase and source fields
\begin{eqnarray}\label{seff}
	& S_{\small\mbox{eff}}(\phi,\alpha)=\frac{1}{2}\int 	dtdt' \gamma^{T}_i{'}[ G^{-1}_{ij}-2\Gamma 	 \delta_{i,j}\delta_{1,j}
	\sigma_x\bar{g}_{\chi}]\gamma^{'}_j,\nonumber\\
	& \bar{g}_{\chi}= e^{i(\chi_c+1/2\sigma_x\chi_q)_t}\bar{g}(t,t')e^{-i(\chi_c+1/2\sigma_x\chi_q)_{t'}}\label{barg}
\end{eqnarray}
Here
$\bar{g}=\left(                                                                                                \begin{array}{cc}                                                                                                  g^R & g^K \\                                                                                                  0 & g^A \\                                                                                                \end{array}\right)$
is the integrated over momentum matrix GF of the lead, while $\Gamma=2\pi \lambda^2 N(0)$ stands for the tunneling width. All the fields are presented in terms of their classical and quantum components as $ \chi_c =(\chi_1+\chi_2)/2$ ; $\chi_q=\chi_1-\chi_2$.

The next step is the integration over the Majorana fermions.
The resulting  generating functional takes the form
\[
  Z(\phi,\alpha)=\exp\left[\frac{1}{2} \mbox{Tr}\ln(G^{-1}-2\Gamma \delta_{i,j}\delta_{1,j}\sigma_x\bar{g}_{\chi})\right].
\]
Integrating over the fields $\phi$ we get the functional to a form which depends only on the source fields
 \begin{eqnarray}
   Z_f(\alpha) &=& \int D(\phi_c, \phi_q)Z(\phi,\alpha)e^{i S_b}, \\
   S_b &=& \int dt \left[\phi_c\left(-\frac{\ddot{\phi_q}}{2E_c}\right)+n_0 \dot{\phi_q}\right].
 \end{eqnarray}
We are looking for a perturbative expression for $\ln Z_f$ to 4th order in tunneling:
\begin{eqnarray}
\ln[Z_f(\alpha)]&=&\ln[Z_f^{(1)}(\alpha)]+\ln[Z_f^{(2)}(\alpha)], \label{z1}
\end{eqnarray}
where the two terms in the right hand side of Eq.~(\ref{z1}) correspond to second and fourth order contributions respectively (see the supplement for details).


{\it Effective capacitance}.---
   The effective capacitance $C_0$ does not coincide with the geometrical capacitance $C$ and is expressed to second order in tunneling through the response function (\ref{cc}) $C_0=e^2 \mbox{Re} K(\omega=0)$ in the the static limit. Generally the real part of $K(t,t')$ takes the form
   \begin{eqnarray}\label{re}
    \mbox{Re} K(tt')&=&\frac{i}{2}\left(\frac{\delta^2 }{\delta \alpha_c (t)\delta \alpha_q (t')}+
    \frac{\delta^2 }{\delta \alpha_q (t)\delta \alpha_c (t')}\right)\ln[Z_f^{(1)}(\alpha)].\nonumber\label{re}
   \end{eqnarray}
Lengthy but straightforward calculations (see Supplement) lead to the following result in the limit of zero temperature ($T\rightarrow 0$) where we concentrate on the nearest two peaks
\begin{eqnarray}
	C_0=e^2\mbox{Re}\,K(\omega=0) = \dfrac{C\Gamma}{4\pi E_c} \sum\limits_{\nu=\pm}{\dfrac{1}{[n_{\nu}+\nu_0/E_c]^2}},\label{rr}
\end{eqnarray}
where $n_{\nu}=1+2n_0 \nu$.

Thus $C_0$ generally depends on the interaction energy $\nu_0$ between the two MBSs. For $\nu_0\neq 0$, the degeneracy points are no longer equally spaced (with a spacing equal to $e$) but they bunch in groups of twos due to alternate spacings $e(1\pm\nu_0/E_c)$ (see Fig.~2 where the entire series of peaks is presented and not only the two resonant ones). See also \cite{ilan1,ilan2} for a quantum Hall analog of this effect. Charge spectroscopy, which is realized by measuring the effective capacitance, can therefore display revealing signatures of the MBSs.

\begin{figure} [!ht]
\centering
\includegraphics[width=0.45 \textwidth]{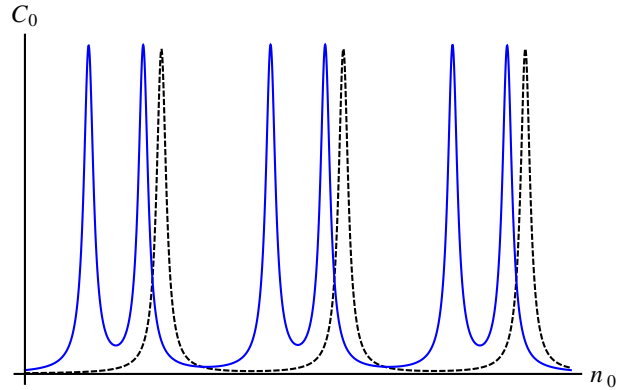}
\caption {\footnotesize The effective capacitance $C_0$ is displayed schematically as function of the background charge $n_0$ for the Majorana Coulomb box problem with spacings that approach $e$ when the Majorana interaction $\nu_0\to 0$ (solid blue line) versus a regular superconductor with equal spacings $2e$ (dashed black line), assuming $\Delta\gg E_c$.}
\label{Fig2}
\end{figure}


{\it Charge relaxation resistance}.---
The imaginary part of density-density correlation function $ \Im K(tt')$ is  obtained  by the variation of $\ln Z_f $ with respect to the source fields. In the limit of small frequency ($\omega\rightarrow 0$) the relevant contribution to $\Im K$ comes from the fourth order term $\ln[Z_f^{(2)}]$ of Eq.~(\ref{z1})
\begin{eqnarray}
   \Im K(tt')&=&\frac{1}{2}\left(\frac{\delta^2 }{\delta \alpha_c (t)\delta \alpha_q (t')}-
   \frac{\delta^2 } {\delta \alpha_q (t)\delta \alpha_c (t')}\right)\ln[Z_f^{(2)}(\alpha)].\nonumber
\end{eqnarray}
To calculate the imaginary part of retarded correlation function we proceed similarly to the case of the real part. Using the second term of Eq.~(\ref{z1}) the frequency dependent imaginary part of response function
$\mbox{Im}\,K(\omega)$ (see supplement) acquires a form
\begin{eqnarray}
 \Im K(\omega)&= &\frac{\omega\Gamma^2}{16\pi}\sum_{\nu\nu'=\pm}\int \frac{d\epsilon_1d\epsilon_3}{(2\pi)^2}\frac{ \theta(\nu\epsilon_1)\theta(\nu'\epsilon_3) G_-(\epsilon_1) G_-(\epsilon_3)}{(E_{\nu}+\nu \epsilon_1)^2(E_{\nu'}+\nu' \epsilon_3)^2},\nonumber
\end{eqnarray}
where $G_{-}=G^R_{11}- G^A_{11}$. Completing the integration over energy variables we obtain the main result of this work
\begin{eqnarray}
 \Im K(\omega)&= &\frac{h\omega}{2E_c^2}\left(\frac{\Gamma}{8\pi E_c}\right)^2\sum_{\nu\nu'=\pm}\frac{ 1 }{\tilde{n}_{\nu}^2 \tilde{n}_{\nu'}^2},\label{ii}
\end{eqnarray}
where $\tilde{n}_{\nu}=n_{\nu}+ \nu_0/E_c $ and we restore the units $h=2\pi$. Using the last formula we see that $\mbox{Im}\,K(\omega)$ satisfies the Korringa-Shiba relation \cite{shiba}
\[\Im K(\omega)=\frac{h}{2}\omega[ReK(0)]^2. \]
Thus we demonstrated the presence of a dissipative charge relaxation resistance in response to the oscillating gate voltage, associated with a unitary resistance value $R_q=h/2e^2$. This result deviates from its expected value when the interaction on the dot is taken to zero ($E_c=0$). The value for the charge relaxation resistance of the Majorana Coulomb box is equal to that of a local impurity level which is described by the Anderson Hamiltonian~\cite{mora2}. Superficially the Majorana Coulomb box in the presence of strong Coulomb blockade mimics the physics of a small quantum dot coupled to a metallic reservoir. The Majorana box discussed here is nevertheless different in several important aspects: (i) the superconducting gap is the largest energy  parameter of the system ($\Delta\gg T_K$). Therefore the critical Fermi liquid hamiltonian (see \cite{mora2}) is not reached and the Kondo effect is excluded; (ii) in addition to regular tunneling, there is also an anomalous (Andreev) term in the tunneling hamiltonian (\ref{hd}) which is absent in the standard tunneling hamiltonian of the Anderson model. The first one describes
the transfer of an electron from the $d$-level to the
normal lead, and the second one the annihilation of a
Cooper pair inside the island accompanied by the creation
of two electrons, one in the $d$-level and one in a
normal electrode; and, (iii) the occupation $\hat{n}_d$ of the $d$-level depends on the fermion parity which imposes a constraint on the Hilbert space (\cite{fu}). We dismantled this constraint (see \cite{egger}) by linking the occupation of the Majorana bound state to the parity of the superconducting condensate.

The principal conclusion that follows from the differences between the Majorana Coulomb box described here and the Anderson model is that for the Majorana case the strong interactions influence the conductance (damping the Andreev tunneling) and, unlike the case without interactions, result in the same value for charge relaxation resistance as that of a quantum dot with a single energy level. This is in stark contrast to the case of the Anderson model where the charge relaxation value is the same independently of interactions.


{\it Resonant tunneling model}.--- Equations ~(\ref{rr}) and (\ref{ii}) have been derived using the Keldysh formalism. It is useful to compare them with an effective resonant tunneling model obtained in~\cite{fu} describing the Majorana box in the strong Coulomb blockade regime. The effective Hamiltonian focuses on two charge states that differ by one electron \cite{fu}. For our setup (Fig.~1) it takes the form
\begin{equation}\label{ham}
    H=H_L+\epsilon_0\left(f^{\dagger}f-\frac{1}{2}\right)+\lambda c^{\dagger}(0)f+\mbox{h.c.},
\end{equation}
where $f^\dagger$ ($f$) are the creation (annihilation) operator for a Dirac fermion occupying the resonant level, $\epsilon_{0}$ is the energy difference between the levels of resonant model and depends on the gate voltage. The density-density correlation function $K(t)$ is expressed in terms of the occupation $n_f=f^\dagger f$ of the resonant level. After we integrate out the reservoir the problem is described by the simple quadratic action
 \begin{equation}\label{fsf}
    S_f=\int dt f^{\dagger} G_f^{-1}f.
\end{equation}
Here the matrix form of the GF $G_f$ is the same as the one for $\bar{g}$ in Eq.~(\ref{barg}) and its entries are: $G_f^R(\epsilon)=G_f^{A*}=(\epsilon-\epsilon_0+i\Gamma /2)^{-1} $, $ G_f^K(\epsilon)=i\Gamma\tanh(\epsilon/2T)
/[(\epsilon-\epsilon_0)^2+\Gamma^2/4]^{-1}$.
With the help of these Green functions we can immediately find the dynamical charge response and effective capacitance (see details in the supplement). In the  low frequency limit we obtain
\begin{eqnarray}
  \Re K(\omega\rightarrow 0) &=&\frac{\Gamma}{2\pi}\frac{1}{\epsilon_0^2+\Gamma^2/4},\label{rrr} \\
  \Im K(\omega)&=& \frac{h\omega}{2} [\Re K(0)]^2. \label{fim}
\end{eqnarray}
By comparing the results of our direct expansion in tunneling, Eqs. (\ref{rr})-(\ref{ii}), with the results obtained using the resonant model, Eqs. (\ref{rrr})-(\ref{fim}), we can find the gate voltage dependence of $\epsilon_0$. To second order in tunneling we get
\begin{equation}\label{eng}
    \epsilon_0=E_c\left[\sum_{\nu=\pm}\frac{1}{2(n_\nu+\nu_0/E_c)^2}\right]^{-1/2}.
\end{equation}
Hence the results derived using the effective resonant model described in~\cite{fu} are fully consistent with our direct perturbation theory approach once the effective parameters are correctly identified.


{\it Conclusions}.--- In this work we used the Keldysh technique to investigate  the dynamical charge response in a system formed by a short topological superconducting mesoscopic wire with two Majorana fermions hosted at the ends of the wire. The wire is coupled on one side to a normal lead and low frequency gate voltage is applied to the system. We calculated the  effective capacitance and charge relaxation resistance and derived a dissipative relaxation resistance associated with a unitary conductance value $R_q=h/(2e^2)$. This is in agreement with Korringa-Shiba formula where the charge relaxation resistance is equal to $R_q$. We demonstrated that this value is also consistent with the effective model for the strong Coulomb blockade limit described in~\cite{fu} and identified its effective parameters. The capacitance $C_0$ generally depends on the interaction between the two Majorana states (see Fig.~2). Both these effects (dissipation and capacitance) can help identify Majorana fermions.
\noindent
\begin{acknowledgments}
We would like to thank  Yu. V. Nazarov  for stimulating discussions and B. Horovitz for valuable remarks. EG acknowledges support from the Israel Science Foundation (grant no.~401/12) and the European Union's Seventh Framework Programme (FP7/2007-2013) under grant agreement no.~303742.
\end{acknowledgments}

\clearpage

\def\up{\uparrow}
\def\down{\downarrow}
\def\ud{\uparrow\downarrow}
\def\du{\downarrow\uparrow}
\def\uu{\up\up}
\def\dd{\down\down}
\def\Tr{\mbox{Tr}\,}
\def\Im{\mbox{Im}}
\def\Re{\mbox{Re}}

\title{Charge Resistance in a Majorana RC circuit. Supporting  Material}

\maketitle

\setcounter{figure}{0}
\setcounter{equation}{0}
\makeatletter
\renewcommand{\thefigure}{S\@arabic\c@figure}
\renewcommand{\theequation}{S\@arabic\c@equation}
\makeatother
\section{Supplementary Information}
\subsection {Perturbation Theory}
The second and fourth order terms in Eq.~(11) of the main text consist also of a functional integration over the phase (see Eq.~(9))
\begin{eqnarray}
  \ln[Z_f^{(1)}(\alpha)] &=&-\Gamma \langle \Tr \Omega(tt)\rangle,  \label{ss1}\\
  \ln[Z_f^{(2)}(\alpha)]  &=& -\Gamma^2\bigg[\frac{1}{2}\Big(\langle \Tr \Omega(tt)\rangle^2-\langle(\Tr \Omega(tt))^2\rangle\Big)+ \label{ss2}\nonumber\\
  &&\langle \Tr [\Omega(tt_1)\Omega(t_1 t)]\rangle\bigg],   \\
\Omega(tt')&=& \int dt_1 G_{11}(tt_1 )\sigma_x \bar{g}_{\chi}(t_1 t').\label{ss3}
\end{eqnarray}
Here the trace includes an integration over the time variables.

{\it Effective capacitance}: To find the effective capacitance we need to calculate the real part of the correlation function $K(\omega\rightarrow 0)$. Explicitly the second order contribution Eq.~(\ref{ss2}) (prior to the functional integration over the phase) acquires a form
  \begin{eqnarray}
    \mbox{Tr}\,\Omega(tt) &=& \frac{1}{4}\int dt dt_1\sum_{\nu,\nu'}e^{i[\chi_c(t_1)- \chi_c(t)+i(\chi_q(t_1)\nu- \chi_q(t)\nu')/2]}\nonumber\\
    &&\times \mbox{Tr}[G_{11}(tt_1)\sigma_x(1+\nu\sigma_x)\bar{g}(t_1t)(1+\nu'\sigma_x)].\quad\label{ss4}
  \end{eqnarray}
Here $\nu,\nu'=\pm1$.
We notice the relation $\chi_{c,q}(t_1)\pm \chi_{c,q}(t)=\phi_{c,q}(t_1)\pm\phi_{c,q}(t)-\int_{0}^{t_1} dt'\alpha_{c,q}(t')\pm\int_{0}^t dt'\alpha_{c,q}(t')$.
Variation with respect to classical source field $\alpha_c$ and integration over $\phi_c$ define a functional $\delta$-function which sets the
value of the quantum field $\phi_q$:
\begin{eqnarray}
 0&=&- \frac{\ddot{\phi}_q (t')} {2E_c} +\delta(t_1-t')- \delta(t-t'),\label{ss5} \\
  \phi_q(t') &=& 2E_c[\theta(t_1-t')(t_1-t')- \theta(t-t')(t-t')].\quad\label{ss6}
\end{eqnarray}
Variation with respect to the quantum source defines $\Re K$ (using Eq.~(\ref{ss6})). The principal contribution to the variation of $\ln Z_f$  comes from $ \nu'=-\nu $ and is equal to
\begin{eqnarray}
  \delta^2 \ln[Z_f^{(1)}]_{\alpha\rightarrow0}(t_f) &=& \frac{\nu}{8}\sum_{\nu=\pm}\int dt dt_1(\theta(t)-\theta(-t_1))\nonumber\\
  && \times e^{-i\nu E_{\nu}(t-t_1)}\check{F}\nonumber\\
   && \times [\theta(t_1-t_f)+\theta(t-t_f)]\theta(t_f),\quad \label{ss7}
\end{eqnarray}
 where $ E_{\nu}=E_c (1-2n_0 \nu)$ and we denote the trace term in Eq.~(\ref{ss4}) as $\check{F}$. We Fourier transform according to $\int dt_f \exp[i\omega t_f]...$, and in the limit of zero frequency the term in the square brackets is integrated to a form $t_1\theta(t_1)+t\theta(t)$.
Direct integrations over $t,t_1$ now result in:
\begin{equation}\label{ss8}
  \mbox{Re}\,K=-\frac{\Gamma}{4}\int \frac{d\epsilon_1}{2\pi}\frac{d\epsilon_2}{2\pi}\sum_{\nu}\frac{\tilde{G}_{11}(\nu,\epsilon_1)
  [g^K-\nu(g^R-g^A)]_{\epsilon_2}}{(E_\nu+\nu(\epsilon_1-
    \epsilon_2))^3},
\end{equation}
where $\tilde{G}_{11}(\nu,\epsilon_1)=[G^K_{11}+\nu (G^R_{11}-G^A_{11})]_{\epsilon_1}$, $g^K(\epsilon)=-i\tanh(\epsilon/2T)$.
Finally, following integration over energy variables and taking the limit $T\to 0$ we arrive at the main result for the effective capacitance presented in  Eqs.~(12)-(13) of the main text.

{\it Charge relaxation resistance}:
Charge relaxation resistance can be extracted from the fourth order term of Eq.~(11). We start our calculations by writing the explicit form of the second term in (\ref{ss2}) :
\begin{eqnarray}
  \Tr [\Omega(tt_1)\Omega(t_1 t)] &=& \frac{1}{16}\int\sum_{\nu_i}\Psi_{(\nu_i)} \Pi_1\Pi_2 dtdt_1dt_2dt_3,\nonumber
\end{eqnarray}
where
\begin{eqnarray}
  \Pi_1&=&\exp\left\{-i[\chi_c(t)+\chi_c(t_1)-\chi_c(t_2)-\chi_c(t_3)]\right\}, \nonumber
\end{eqnarray}
and
\begin{eqnarray}
  \Pi_2&=& \exp\left[\frac{i}{2}(\nu_1\chi_q(t_2)-\nu_2\chi_q(t_1)+\nu_3\chi_q(t_3)-\nu_4\chi_q(t)\right].\nonumber
\end{eqnarray}
Here the integration is over all time variables and summation refers to all numbers $\nu_i=\pm 1$ ($i=1,2,3,4$).
The phases $\chi$ are the sum of charge and source fields, and
\begin{eqnarray}
\Psi_{(\nu_i)} &=& \mbox{Tr}[G_{11}(\epsilon_1)\sigma_x(1+\nu_1\sigma_x)\bar{g}(\epsilon_2)(1+\nu_2\sigma_x)\nonumber\\
     && G_{11}(\epsilon_3)\sigma_x(1+\nu_3\sigma_x)\bar{g}(\epsilon_4)(1+\nu_4\sigma_x)].\label{sp9}
\end{eqnarray}
The above $\Psi$ is analogous to the function $\hat{F}$ introduced before.
Performing variations with respect to the quantum and classical source fields  we arrive at the following expression for $\Im K(\omega)$
\begin{eqnarray}
  \Im K(\omega) &=& \frac{\Gamma^2}{32} \Im \int \sum_{\nu_i} [\theta(t)+ \theta(t_1)-\theta(t_2)-\theta(t_3)]\label{ss9} \nonumber\\
  && \times\Psi_{(\nu_i)} e^{i E_c \left[D+2n_0(t+t_1-t_2-t_3)\right]}r(\omega),
\end{eqnarray}
where the terms in square brackets originate from the variation with respect to the classical source, while the function
 \begin{eqnarray}
     r(\omega) &=&\frac{1}{2i\omega}(-\nu_1 e^{i\omega t_2}+\nu_2e^{i\omega t_1}-\nu_3e^{i\omega t_3}+\nu_4 e^{i\omega t}),\nonumber
   \end{eqnarray}
appears as a result of the variation with respect to the quantum source $\alpha_q$ and is dependent on the path followed by the electron. The functional integration on the classical component of the charge field $\phi_c$ with the action given in Eq.~(10), defines the function $D$ in the exponent of Eq.~(\ref{ss9}). To calculate $D$ we use the solution for the quantum field $\phi_q$ which is found from an equation similar to Eq.~(\ref{ss5}) and results from integration over $\phi_c$,
\begin{eqnarray}\label{ss10}
 0&=&- \frac{\ddot{\phi}_q (t')} {2E_c} +\delta(t_2-t')- \delta(t_1-t')+\nonumber\\
 &&\delta(t_3-t')- \delta(t-t'), \nonumber\\
    \phi_q(t')& =& 2E_c[\theta(t_2-t')(t_2-t')- \theta(t_1-t')(t_1-t')+\nonumber\\
    &&\theta(t_3-t')(t_3-t')- \theta(t-t')(t-t')].
\end{eqnarray}
Finally, replacing the charge quantum phase by its solution (\ref{ss10}) we find the function $D$ appearing in the exponent of (\ref{ss9}). For a general path followed by the electron, $D$ takes a form
\begin{eqnarray}
  D &=& (t_2-t_1)[\nu_1\theta(t_1-t_2)-\nu_2\theta(t_2-t_1)-\epsilon_2]+\nonumber \\
  &&(t_3-t_2)[ \nu_1\theta(t_3-t_2)-\nu_3\theta(t_2-t_3)]+\nonumber\\
  &&(t_2-t)[\nu_1\theta(t-t_2)-\nu_4\theta(t_2-t)+\epsilon_1]+\nonumber\\
  &&(t_1-t_3)[\nu_2\theta(t_3-t_1)-\nu_3\theta(t_1-t_3)-\epsilon_3]+\nonumber\\
  &&(t-t_1)[\nu_2\theta(t-t_1)-\nu_4\theta(t_1-t)]\nonumber+\nonumber\\
  &&(t_3-t)[\nu_3\theta(t-t_3)-\nu_4\theta(t_3-t)-\epsilon_4].\label{D}\nonumber
\end{eqnarray}
In the definition of $D$ we have included the energy variables (in units of $E_c$)  which come from the Fourier images of the Green functions [see $\Psi_{(\nu_i)}$ in Eq.~(\ref{sp9})]. Thus $D$ defines the path in time which the electron follows. We identify each path by four numbers as $(\nu_1 \nu_2 \nu_3 \nu_4)$. The electron-hole excitations in the normal metal lead provide the main contribution to the dissipation. This fixes $4$ relevant path configurations: $(a)(-111-1),(b)(1-1-11),(c)(-1-1-1-1),(d)(1111)$. For these paths the trace in Eq.~(\ref{sp9}) (taking the relevant part which contributes to $\Im K$) acquires the following form at $T\rightarrow0$
\begin{eqnarray}
  \Psi_{(a),(b)}&=& 4\theta(\pm\epsilon_2)\theta(\mp\epsilon_4)[G^K_{11}(\epsilon_1)G^K_{11}(\epsilon_3)-G_+(\epsilon_1)G_+(\epsilon_3)],\nonumber\\
  \Psi_{(c)}&=&\Psi_{(d)}=\sum_{\eta=\pm}\theta(\eta\epsilon_2)\theta(-\eta\epsilon_4)[G^K_{11}(\epsilon_1)G^K_{11}(\epsilon_3)+\nonumber\\
  &&G_+(\epsilon_1)G_+(\epsilon_3)].\label{ss11}
\end{eqnarray}
Here and below $G_{\pm}=G^R_{11}\pm G^A_{11}$. In Eq.~(\ref{ss11}) we kept only terms which have poles in the particle-hole polarization $(\epsilon_2-\epsilon_4\pm \omega)^{-1}$. For the four relevant paths $(a),(b),(c),(d)$ we also calculate the function $D$ and find the corresponding contribution to $\Im K$ by performing the time integrations in Eq.~(\ref{ss9}). The time integrations fix the signs of energy variables and result in a factor
\begin{eqnarray}
  &&\sum_{\nu\nu'=\pm 1}\frac{i\omega (-1)^{1+(\nu+\nu')/2}\theta(\nu\epsilon_1)\theta(\nu'\epsilon_3)}{(E_{\nu}+\nu \epsilon_1)^2(E_{\nu'}+\nu' \epsilon_3)^2}.\label{factor}
\end{eqnarray}
Combining the GF with Eq.~(\ref{factor}) and collecting all contributions we arrive at Eq.~(14) for $\Im K(\omega)$ of the main text.

\subsection{Effective resonant model}
An effective resonant model describing the physics of the Majorana box in the strong Coulomb blockade regime is given by the Hamiltonian in Eq.~(14). After integrations on reservoir variables the problem is reduced to the pure quadratic action Eq.~(15).
 This action is taken on the Keldysh space (either directly or rotated by the matrix $L$ of the main text). In direct space  the correlation function is presented by time ordering products of $n_f=f^{\dagger}f$ operators
 \begin{eqnarray}
   K^R(t-t') &=& i\langle T[n_f^{(1)}(t) n_f^{(1)}(t')-n_f^{(1)}(t) n_f^{(2)}(t')\rangle,\nonumber\\
  \Re K^R(t-t') &=& \frac{i}{2}\langle T[n_f^{(1)}(t) n_f^{(1)}(t')-n_f^{(2)}(t) n_f^{(2)}(t')\rangle,\nonumber \\
   \Im K^R(t-t') &=& \frac{1}{2}\langle T[n_f^{(2)}(t) n_f^{(1)}(t')-n_f^{(1)}(t) n_f^{(2)}(t')\rangle,\nonumber
 \end{eqnarray}
where the numbers in brackets denote the Keldysh indices on the time contour. Using Wick theorem we can contract the $f$-operators and express the above values in terms of the associated Green functions $G_f$
\begin{eqnarray}
 \Re K^R(\omega\rightarrow 0) &=& \frac{i}{2}\int\frac{d\epsilon}{2\pi}\tanh
 \left(\frac{\epsilon}{2T}
 \right)
  G_{f-}(\epsilon)G_{f+}(\epsilon),\nonumber \\
\Im K^R(\omega) &=& \frac{1}{4}\int\frac{d\epsilon}{2\pi}\left[\tanh\left(\frac{\epsilon+\omega}{2T}\right)-\tanh\left(\frac{\epsilon}{2T}\right)\right]\nonumber\\
   &&\times G_{f-}(\epsilon+\omega)G_{f-}(\epsilon),\nonumber
 \end{eqnarray}
where the form of the Green functions $G_{f\pm}=G^R_f\pm G^A_f$ and $G_f$ is presented in the main text. Direct integrations over the energy variables result in Eqs.~(17),(18) of the main text.
\end{document}